\documentclass[floatfix,twocolumn,showpacs,prl,superscriptaddress]{revtex4}
\usepackage{graphicx}
\usepackage{bm}
\usepackage{dcolumn}

\begin{document}

\title{Rashba Effect at Magnetic Metal Surfaces}
\author{O. Krupin}
\affiliation{Institut f\"ur Experimentalphysik, Freie
Universit\"at Berlin, Germany}
\author{G. Bihlmayer}
\affiliation{Institut f\"ur Festk\"orperforschung, 
Forschungszentrum J\"ulich, Germany}
\author{K. Starke*}
\affiliation{Institut f\"ur Experimentalphysik, Freie
Universit\"at Berlin, Germany}
\author{S. Gorovikov}
\affiliation{MAX-Lab, Lund University, Sweden}
\author{J. E. Prieto}
\affiliation{Institut f\"ur Experimentalphysik, Freie 
Universit\"at Berlin, Germany}
\author{K. D\"obrich}
\affiliation{Institut f\"ur Experimentalphysik, Freie
Universit\"at Berlin, Germany}
\author{S. Bl\"ugel}
\affiliation{Institut f\"ur Festk\"orperforschung, 
Forschungszentrum J\"ulich, Germany}
\author{G. Kaindl}
\affiliation{Institut f\"ur Experimentalphysik, Freie
Universit\"at Berlin, Germany}

\date{\today}

\begin{abstract}
We give experimental and theoretical evidence of the Rashba 
effect at the magnetic rare-earth metal surface Gd(0001). 
The  Rashba effect is substantially 
enhanced and the Rashba parameter changes its sign
when a metal-oxide surface layer is formed.
%
The experimental observations are quantitatively described by
{\em ab initio} calculations that
give a detailed account of the near-surface charge density gradients
causing the Rashba effect. Since the sign of the Rashba splitting 
depends on the magnetization direction, the findings open up
new opportunities for the study of surface and interface magnetism.
\end{abstract}

\pacs{71.70.Ej, 72.25.-b, 73.20.-r, 85.75.-d}

\maketitle

 
A key issue in condensed-matter research aiming at
future spintronic devices~\cite{Wolf:01.1} is to control 
and manipulate the electron spin in a two-dimensional electron
gas (2DEG) of semiconductor systems without the need 
of applying an external magnetic field.
Rashba had realized early on~\cite{Rashba:60.1} that 
this can be achieved by
an electric field which
acts as a magnetic field in the rest frame of a moving 
electron. The 
interaction between the spin $\mathbf{s}$ of a moving electron of
momentum $\hbar\mathbf{k}$ 
with an electric field oriented along 
the $z$-axis $\mathbf{e}_{z}$
is described by the Rashba Hamiltonian

\vspace*{-8pt}
\begin{equation}
{\cal H}_{R} = \alpha_{R} \, 
 (\mathbf{e}_z  \times {\mathbf{k}})\cdot \mathbf{s}\, .
\label{eq:rashba}
\end{equation}

\noindent
The Rashba parameter $\alpha_{R}$ is proportional to the
electric field and depends on the effective, material-dependent spin-orbit coupling (SOC) strength. 
In nonmagnetic systems the Rashba effect
lifts the spin-degeneracy of the energy dispersion $\epsilon(\mathbf{k})$ 
of an electronic state,
and the energy difference 
between $\epsilon_{\uparrow}(\mathbf{k})$ and $\epsilon_{\downarrow}(\mathbf{k})$ 
is called Rashba splitting 
$\Delta \epsilon(\mathbf{k}) = \alpha_R \left|\mathbf{k}\right|$.
Even though spintronic research currently focuses on 
spin-polarized electrons 
in semiconductors~\cite{Zhu:01.1,jwd03}, it 
is important 
to explore the Rashba effect in other material classes as well.

A necessary condition for the Rashba effect 
to occur
is the absence of inversion symmetry and, while in the proposed FET-type 
{\em spin transistor}~\cite{Datta:90.1} 
a gate voltage must be applied to break inversion symmetry of the 2DEG,
this condition is naturally fulfilled by the
structural inversion asymmetry (SIA)
existing at any crystal surface or interface.
Owing to SIA, electrons in a two-dimensional surface or interface state 
experience an effective crystal potential gradient perpendicular to
their plane of propagation, hereby optimizing 
$(\mathbf{e}_z  \times {\mathbf{k}})$
in Eq.~(\ref{eq:rashba}).
One should expect that the Rashba effect is 
a general surface and interface phenomenon, but up to now 
Rashba splittings 
have only been observed for surface states at
Au(111)~\cite{LaShell:96.1,Reinert:01.1} and W(110)~\cite{Rotenberg:98.1,Rotenberg:02.1}.
Recently relativistic density functional theory (DFT) calculations 
were able to reproduce the observed 
splitting of the Au $sp$-like surface state~\cite{Nicolay:01.1}
and the analogy to a 2DEG has been pointed out~\cite{Henk:03.1}. 
Yet, it is still a challenging task to 
give a physical picture of the Rashba effect from 
the electronic structure point of view.

This Letter
presents the first experimental and theoretical evidence of a 
Rashba splitting of 
{\em exchange-split} two-dimensional electron states.
Using the surface state of ferromagnetic Gd metal as example 
we report on the novel finding 
of a k-dependent contribution to the binding energy of this state that
changes sign upon magnetization reversal.
It is further demonstrated that the Rashba effect is enhanced 
upon formation of a surface oxide layer. 
The experimental observations 
are quantitatively described by {\em ab initio} calculations
showing that the enhancement 
is caused by a substantial change of charge-density gradients
at the interface between the surface-oxide layer and 
the bulk-like metal, which
leads to an admixture of $p$ character to the 
$d$-derived two-dimensional state.
%
%
%

In particular, we show that the Rashba effect 
in magnetic systems bears interesting consequences
when the spin-degeneracy of $\epsilon(\mathbf{k})$ is already 
lifted by an exchange splitting $\Delta E_{ex}$, separating 
majority ($\uparrow$) and minority ($\downarrow$) electrons.
The dispersion
of a state subject to the Rashba effect can then be written as 
$\epsilon_{\downarrow(\uparrow)}(\mathbf{k}) =%
\epsilon(\mathbf{k}) \pm \frac{1}{2}\,\Delta E_{ex} \pm \frac{1}{2}\,
\alpha_R \left| \mathbf{k} \right|$
and the Rashba splitting 
$ \Delta \epsilon_{\downarrow(\uparrow)}(\mathbf{k})= 
\epsilon_{\downarrow(\uparrow)}(\mathbf{k},\mathbf{M})
                - \epsilon_{\downarrow(\uparrow)}(\mathbf{k},-\mathbf{M})
		= \pm \alpha_R \left|\mathbf{k}\right|$
can be observed separately for majority and minority states
by two subsequent measurements with opposite sample magnetization,
$\mathbf{M}$ and $-\mathbf{M}$. 
Since in many cases the Rashba splitting 
is non-zero only for surface and interface states, the observation 
of this quantity opens a new 
and powerful spectroscopic path 
to discriminate surface and interface magnetism from bulk magnetism.

%
%
%

For the present investigation we 
chose the (0001) surface of ferromagnetic hcp Gd metal 
where the exchange interaction is known to
separate majority and minority branches of the $d$-derived surface state
by several hundred meV~\cite{Weschke:96.1}. 
With this substantial exchange splitting 
a limitation in previous studies of non-magnetic systems is avoided, 
where the Rashba splitting of a state, in order to be 
resolved, must be larger than its intrinsic (lifetime) width~\cite{Reinert:01.1}. 
The remanent {\em in-plane} magnetization of thin Gd(0001) 
films~\cite{Berger:94.1} means that the spin-quantization axis $\mathbf{s}$ is orthogonal to the surface electric 
field, providing an advantageous geometry for observing a
Rashba splitting, see Eq.~(1).
The $p(1\times 1)$O/Gd(0001) surface oxide
allows furthermore to study an exchange-split pair of two-dimensional states~\cite{Schussler:99.1} that are both occupied.

\begin{figure}[tb]
\includegraphics[angle=0,width=80mm]{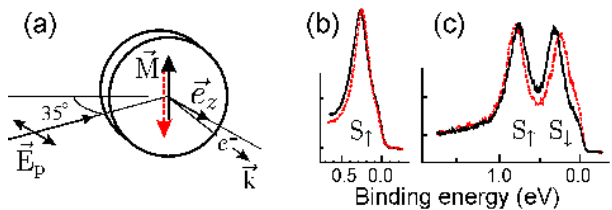}
\caption{(color online)
(a) Experimental geometry. The sample magnetization $\mathbf{M}$
is perpendicular to the plane defined by the surface normal
$\mathbf{e}_z$ and the electron momentum $\mathbf{k}$.
(b,c) Angle-resolved PE spectra at $6^{\circ}$ electron emission angle 
w.r.t.\ the surface normal (excited with p-polarized radiation). 
(b) Gd(0001) majority-spin surface state 
${\rm S_{\uparrow}}$ (36~eV photon energy).
(c) Exchange-split p$(1\times1)$O/Gd(0001) 
interface states ${\rm S_{\uparrow}}$ and ${\rm S_{\downarrow}}$
(45~eV photon energy).
The peak positions change upon magnetization reversal
owing to the Rashba effect.
}
\label{fig1}
\end{figure}

Angle-resolved photoemission (PE)
experiments were performed using display-type
electron analyzers at the I-311 undulator beamline at
MAX-Lab, Lund, and at the BUS beamline at BESSY,~Berlin. 
In the experiments, the energy resolution was set to 30~meV
and the angular resolution to $0.5^{\circ}$.
Gd(0001) films were prepared $in~situ$ by metal vapor deposition on 
a W(110) single crystal mounted 
on a liquid-He flow cryostat.
The film thickness was 10~nm as determined by a quartz
balance. During evaporation the pressure in the UHV chamber 
rose from $5\times10^{-11}$~mbar to $5\times10^{-10}$~mbar.
The surface monoxide layer was
prepared according to the recipe given in Ref.~\cite{Schussler:99.1}
with 2~Langmuir oxygen exposure; it was
controlled {\em in~situ} by PE and LEED. The PE
spectra were recorded from remanently magnetized samples in
the experimental geometry shown in Fig.~\ref{fig1}(a). 


\begin{figure}[tb]
\includegraphics[angle=0,width=70mm]{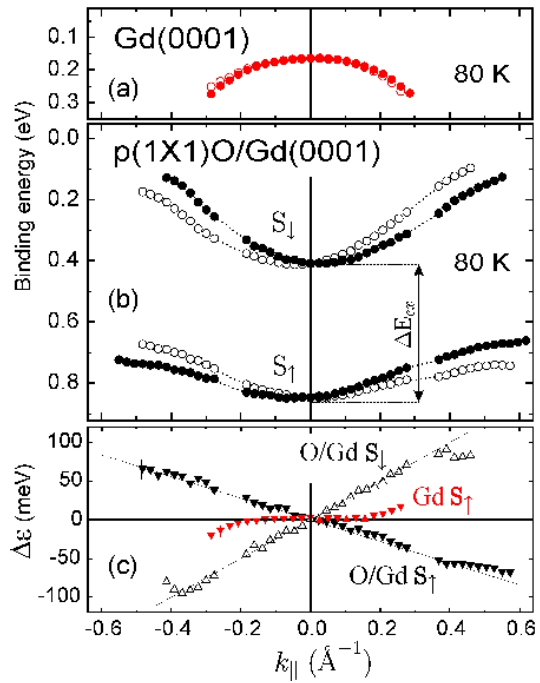}
\caption{(color online)
(a) 
Experimental dispersion of the majority spin Gd(0001) surface state 
in the ${\overline\Gamma}-\overline{\mathrm M}$ azimuth; 
solid (red) and open symbols correspond to opposite 
magnetization directions, see Fig.~1(a).
(b)
${\rm S_{\uparrow}}$ (majority spin) and ${\rm S_{\downarrow}}$ 
(minority spin) interface states 
of $p(1\times 1)$O/Gd(0001) in the 
${\overline\Gamma}-\overline{\mathrm K}$ azi\-muth,
exchange split by $\Delta E_{ex}=450$~meV at ${\overline\Gamma}$.
(c)
Rashba splitting $\Delta \epsilon$ derived from the
data in (a) (indicated by Gd-${\rm S_{\uparrow}}$) and (b) 
(indicated by O/Gd-${\rm S_{\uparrow}}$ and O/Gd-${\rm S_{\downarrow}}$). 
} 
\label{fig2}
\end{figure}

For ferromagnetic Gd metal only the majority component 
${\rm S_{\uparrow}}$ of the $d$-derived Gd(0001) surface state is occupied~\cite{Weschke:96.1},
as is shown by the angle-resolved PE spectrum in Fig.~1(b). 
At off-normal emission angles, the 
${\rm S_{\uparrow}}$ peak position shifts significantly upon
magnetization reversal. Derived from experimental spectra,
the dispersion near the center of the surface Brillouin zone 
along the $\overline\Gamma$-$\overline\mathrm{M}$ direction
is presented in Fig.~\ref{fig2}(a);
filled and open data points distinguish the two branches 
measured for opposite magnetization 
directions (cf.\ Fig.~1(a)). 
The branches are shifted 
symmetrically 
with respect to $\mathbf{k}=0$,
in agreement with the triple-vector product of Eq.~(1).
The energy difference $\Delta \epsilon (\mathbf{k}_{\|})$ 
is plotted in Fig.~2(c). It is identified as Rashba splitting 
of the Gd(0001) surface state;
it remains rather small ($< 25$~meV) in the $\mathbf{k}_{\|}$ 
range of ${\rm \pm 0.3~\AA^{-1}}$.

The $p(1\times 1)$O/Gd(0001) surface oxide
exhibits an exchange-split pair of surface bands~\cite{Schussler:99.1}
that are both occupied, shown 
as ${\rm S_{\uparrow}}$ and ${\rm S_{\downarrow}}$
in Fig.~\ref{fig1}(c). 
For nonzero $\mathbf{k}_{\|}$ (off-normal emission),
the peak positions clearly change into opposite directions
when the magnetization is reversed.
Their (upward) dispersions are shown in Fig.~2(b): 
they change symmetrically with respect to the Brillouin zone center,
strongly supporting our interpretation as Rashba splitting.
The energy splitting $\Delta \epsilon (\mathbf{k}_{\|})$ 
is plotted separately for S$_\uparrow$ and S$_\downarrow$ in Fig.~2(c);
it is substantially larger (about three times at 
${\rm \left| \mathbf{k}_{\|}\right|= 0.3~\AA^{-1}}$) than 
the Rashba splitting of the majority surface state on Gd(0001).
Moreover, the Rashba parameters 
for the oxide-covered and the clean metal surface 
are found to have opposite signs, see Fig.~2(c). --
We note in passing that for $p(1\times 1)$O/Gd(0001)
also the modulus of the slopes, 
$\left| \alpha_R \right|$, is different
for ${\rm S_{\uparrow}}$ and ${\rm S_{\downarrow}}$.

\begin{figure}[tb]
\includegraphics[angle=0,width=80mm]{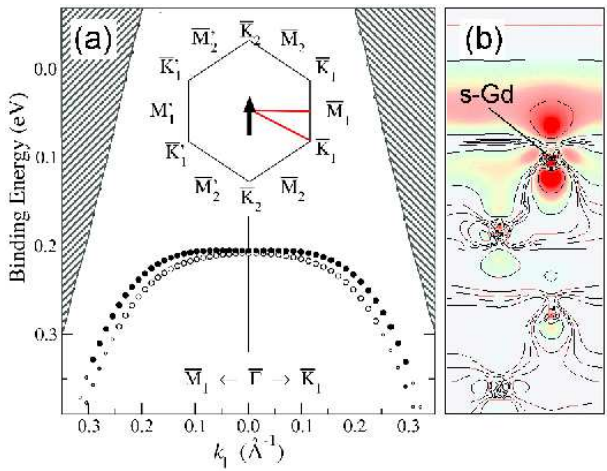}
\caption{(color online) (a) 
Calculated majority-spin surface-state dispersions of 
Gd(0001) along two different high-symmetry directions 
(shown as red lines) of the 
magnetic surface Brillouin zone, given as inset.
Hatched areas indicate bulk band regions.
For symmetry labels, see text.
(b) 
Charge density distribution (isolines on log scale) for ${\rm S_\uparrow}$
at $\overline{\Gamma}$ in a plane perpendicular to the surface.
The linear color scale ranges from red (high charge density), 
yellow (medium), green (low), blue (very low) to white (negligible).
} \label{fig:Gd_Bs_th}
\end{figure}

The {\em ab initio} calculations were performed using DFT in the local
density approximation (LDA) employing 
the form of Moruzzi~et~al.~\cite{Moruzzi:78.1}.
We use the full-potential linearized 
augmented-plane-wave method in film geometry~\cite{Wimmer:81.1,www-fleur},
with SOC included self-consistently according to Ref.~\cite{Li:90.1}.
For a proper description of the $4f$ electrons we applied the 
LDA+U method~\cite{Kurz:02.1}.
The Gd surface was simulated by a structurally relaxed 
10-layer film embedded in two semi-infinite vacua.  
A plane-wave cutoff of $k_{\rm max} = 3.8$~(a.u.)$^{-1}$ 
was used, 
and the irreducible part of the two-dimensional Brillouin-zone (BZ) was 
sampled at 21 special $\mathbf{k}_{\|}$ points
(36 $\mathbf{k}_{\|}$ points for calculations with SOC included).

In the 
calculations with SOC included, 
we took into account that the in-plane magnetization 
lowers the $p3m1$ symmetry of hcp Gd$(0001)$ to $cm$ symmetry. 
One can still label the high
symmetry points of the BZ as 
$\overline{\mathrm M}$ and $\overline{\mathrm K}$, 
but with subscripts to distinguish between inequivalent points, 
see Fig.~\ref{fig:Gd_Bs_th}(a). 
Points that are related by an inversion center
are primed to indicate that the bandstructure is equivalent to the unprimed
$\mathbf{k}_{\|}$-points if one interchanges spin-up and spin-down bands.
The magnetization was assumed to point in the direction
of the in-plane nearest neighbors. 
The size of the Rashba splitting decreases from 
the $\overline{\Gamma}$-$\overline{\mathrm M}_1$ direction
to  $\overline{\Gamma}$-$\overline{\mathrm K}_1$ and further to
$\overline{\Gamma}$-$\overline{\mathrm M}_2$, and is zero in 
$\overline{\Gamma}$-$\overline{\mathrm K}_2$ direction.
%
%

\begin{figure}[tb]
\includegraphics[angle=0,width=80mm]{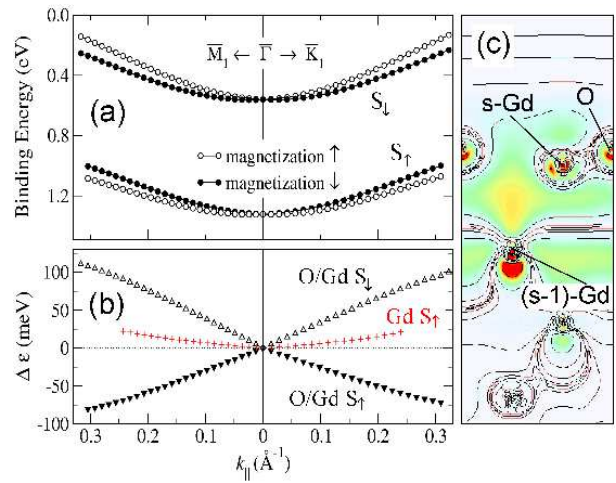}
\caption{(color online) (a) Calculated surface-state dispersions of 
$p(1\times 1){\rm O/Gd(0001)}$ along the 
high symmetry directions 
$\overline{\Gamma}$-$\overline{\mathrm M}_1$ 
(left side) and
$\overline{\Gamma}$-$\overline{\mathrm K}_1$
(right side)
of the surface BZ, see red lines in Fig.~3(a); 
(b) Rashba splittings, $\Delta \epsilon$, for 
majority and minority bands (triangles) and for the  
clean Gd surface (red crosses) given for comparison. 
(c) Charge density distribution
for ${\rm S_\downarrow}$ at $\overline{\Gamma}$;
a similar plot has been obtained for ${\rm S_\uparrow}$.
The linear color code is like in Fig.~3.} 
\label{fig:OGd_Bs_th}
\end{figure}

The theoretical results for $p(1\times 1)$O/Gd(0001) are 
presented in Fig.~4.
In the calculation, the
O atoms were 'adsorbed' on the fcc site and its relaxed
position was found 0.78~\AA\ above the plane 
of Gd surface atoms, s-Gd. Upon O adsorption, the s-Gd layer
shows a strong (18\%) outward relaxation, while the position of
the inner layers remains almost unperturbed. 
Very similar results were obtained for O adsorbed on 
the energetically slightly less favorable hcp site. 
In Fig.~4(a), the two occupied states disperse upwards 
(positive effective mass)
in good agreement with experiment, see Fig.~\ref{fig2}.
(The exchange splitting 
is somewhat larger than the 
experimental value, an overestimation that had been noted before
for clean Gd(0001)~\cite{Kurz:02.1}.)
The calculated Rashba splitting $\Delta \epsilon$ of majority-spin 
and minority-spin bands is shown in Fig.~4(b); it is about
{\em three times larger} than that of the Gd(0001) metal surface, 
and $\alpha_R$ is found to be of opposite sign, 
in excellent agreement with experiment.

With the results of the present {\em ab initio} calculations, we can
address the question why the oxide layer on 
Gd(0001) causes an enhanced Rashba splitting.
Figs.~3(b) and~4(c) display the charge density 
distributions of the two-dimensional states 
at $\overline{\Gamma}$ for the two systems, Gd metal and O/Gd, 
in a plane perpendicular to the surface. 
In addition to the conventional isolines (log scale),
the surface-state charge densities are also 
given on a linear scale (color) for better visibility.
When comparing the profiles it becomes obvious that adsorption 
of the electronegative O atom
changes the charge density distribution of the 
entire near-surface region. 
While the surface state of the metal surface, cf.\ Fig.~3(b),
resides almost exclusively (to $\sim 89$\%) in the top surface
layer s-Gd,
the two-dimensional state in $p(1\times 1)$O/Gd is 
distributed over both the s-Gd and (${\rm s-1}$)-Gd
layer, see Fig.~4(c). Hence, given the close vicinity of 
O and s-Gd layer, this state may be conceived as
{\em interface state} located between a 
$p(1\times 1)$O/Gd surface monoxide layer
and bulk-like Gd metal.

At Gd(0001) the surface-state charge is located quite 
symmetrically below and above the s-Gd plane,
cf.\ Fig.~3(b), corresponding to a 
small (positive) charge density gradient along
the surface normal. 
The small charge density gradient directly indicates that there 
is a small electric field 
in the s-Gd layer of Gd(0001), which gives rise to the small Rashba splitting, shown in Figs.~2(c) and 4(b).
At $p(1\times 1)$O/Gd(0001), by contrast, 
the rather asymmetric charge distribution of the interface state
in the (${\rm s-1}$)-Gd layer directly indicates the presence of a 
high effective electric field at this layer.
The charge gradient is negative, 
i.e.\ opposite to the uncovered metal surface. We are led to conclude
that it is this reversed effective electric field 
in the (${\rm s-1}$)-Gd layer that causes the opposite sign of the Rashba 
parameter of the interface state
in $p(1\times 1)$O/Gd(0001) as compared with Gd(0001).

Furthermore, the calculations reveal that the surface oxide 
formation is accompanied by a pronounced change in {\em orbital character}
of the two-dimensional state. 
In the s-Gd layer it changes upon oxidation from almost 
exclusively $d_{z^2}$-like ($d:p$-ratio $\approx 8:1$, 
integrated over the muffin-tin sphere) to predominantly 
$s$-like with an admixture of other orbital components
($d:p:s \approx 3:2:11$).
In the (${\rm s-1}$)-Gd layer of $p(1\times 1)$O/Gd 
the state remains $d_{z^2}$-like yet with a substantial
$p_{z}$-admixture ($d:p \approx 5:1$).
Moreover, in order to identify the relative contributions 
of the individual layers to the Rashba splitting,
we calculated hypothetical values of the splitting
with SOC set to zero for all other layers.
Again there is a striking difference of the two systems: 
at Gd(0001), the s-Gd layer provides
by far the main contribution ($\sim 90$\%) to the Rashba splitting;
yet in the surface oxide system
it is the (${\rm s-1}$)-Gd layer that accounts for over 70\%
of the splitting, whereas the
s-Gd layer of $p(1\times 1)$O/Gd cannot contribute 
owing to its prevailing s~character (see above).

In the light of these results we arrive at the following 
physical picture of the Rashba effect.
At the Gd(0001) metal surface SIA leads to 
a small but significant spill-out of the $d$-derived surface state,
yet the charge gradient is small indicated
by the relatively small admixture of $p_z$-character 
(antisymmetric w.r.t.\ the surface plane)
to the $d_{z^2}$-derived state.
With the epitaxial $p(1\times 1)$O/Gd surface oxide layer 
present, the electronegative O attracts charge from the s-Gd atom 
so that strong {\em charge-density gradients} arise not only in the 
s-Gd layer but also in the sub-surface layer.
As a consequence,
the two-dimensional state changes in spatial distribution 
(becoming an interface state) and in orbital character.

In summary, we have demonstrated that in magnetic systems 
with sufficiently large exchange splitting, 
${\rm \Delta E_{ex} \gg \Delta \epsilon_{\downarrow(\uparrow)}}$, 
i.e.\ when majority ($\uparrow$) and minority spin ($\downarrow$) 
electronic states are well separated,
even small Rashba splittings can be determined by two 
subsequent measurements with opposite sample magnetizations. 
Since one can expect an analogous behavior 
for other magnetic materials, a
measurement of the Rashba splitting opens up a new and powerfull 
way
to discriminate surface and interface magnetism from bulk magnetism. 
The present discovery of a particularly large Rashba effect
at an interface between a
two-dimensional metal oxide and a magnetic metal
should stimulate future research 
towards a potential use of such
interfaces for spintronic devices.
%
%
%
%
-- 
Moreover, based on the present findings it appears to be particularly interesting to study the evolution of an exchange-split two-dimensional state into a laterally confined quantum-well state of a magnetic nanostructure where the elastic reflection of the state $|\mathbf{k_\|}\rangle$ is suppressed owing to the Rashba effect, since the reflected state $ |-\mathbf{k_\|}\rangle$ is energetically not accessible.

We gratefully acknowledge expert experimental support by J.\ Andersen 
(MAX-Lab) and R. P\"uttner (BUS beamline, BESSY).
The work in Berlin was supported by BMBF, Contract 05 KS1 KEC/2,
and DFG (SfB-290).

\vspace*{4pt}
$^*$Corresponding author: {\em starke@physik.fu-berlin.de}

\vspace*{-8pt}
\bibliography{rashba}

\begin{thebibliography}{19}
\expandafter\ifx\csname natexlab\endcsname\relax\def\natexlab#1{#1}\fi
\expandafter\ifx\csname bibnamefont\endcsname\relax
  \def\bibnamefont#1{#1}\fi
\expandafter\ifx\csname bibfnamefont\endcsname\relax
  \def\bibfnamefont#1{#1}\fi
\expandafter\ifx\csname citenamefont\endcsname\relax
  \def\citenamefont#1{#1}\fi
\expandafter\ifx\csname url\endcsname\relax
  \def\url#1{\texttt{#1}}\fi
\expandafter\ifx\csname urlprefix\endcsname\relax\def\urlprefix{URL }\fi
\providecommand{\bibinfo}[2]{#2}
\providecommand{\eprint}[2][]{\url{#2}}

\bibitem[{\citenamefont{Wolf et~al.}(2001)\citenamefont{Wolf, Awschalom,
  Buhrman, Daughton, von Molnar, M.~L.~Roukes, and Treger}}]{Wolf:01.1}
\bibinfo{author}{\bibfnamefont{S.~A.} \bibnamefont{Wolf}},
  \bibinfo{author}{\bibfnamefont{D.~D.} \bibnamefont{Awschalom}},
  \bibinfo{author}{\bibfnamefont{R.~A.} \bibnamefont{Buhrman}},
  \bibinfo{author}{\bibfnamefont{J.~M.} \bibnamefont{Daughton}},
  \bibinfo{author}{\bibfnamefont{S.}~\bibnamefont{von Molnar}},
  \bibinfo{author}{\bibfnamefont{A.~Y.~C.} \bibnamefont{M.~L.~Roukes}},
  \bibnamefont{and} \bibinfo{author}{\bibfnamefont{D.~M.}
  \bibnamefont{Treger}}, \bibinfo{journal}{Science}
  \textbf{\bibinfo{volume}{294}}, \bibinfo{pages}{1488} (\bibinfo{year}{2001}).

\bibitem[{\citenamefont{Rashba}(1960)}]{Rashba:60.1}
\bibinfo{author}{\bibfnamefont{E.~I.} \bibnamefont{Rashba}},
  \bibinfo{journal}{Sov.\ Phys.\ Solid State} \textbf{\bibinfo{volume}{2}},
  \bibinfo{pages}{1109} (\bibinfo{year}{1960}).

\bibitem[{\citenamefont{Zhu et~al.}(2001)\citenamefont{Zhu, Ramsteiner,
  Kostial, Wassermeier, Sch\"onherr, and Ploog}}]{Zhu:01.1}
\bibinfo{author}{\bibfnamefont{H.~J.} \bibnamefont{Zhu}},
  \bibinfo{author}{\bibfnamefont{M.}~\bibnamefont{Ramsteiner}},
  \bibinfo{author}{\bibfnamefont{H.}~\bibnamefont{Kostial}},
  \bibinfo{author}{\bibfnamefont{M.}~\bibnamefont{Wassermeier}},
  \bibinfo{author}{\bibfnamefont{H.~P.} \bibnamefont{Sch\"onherr}},
  \bibnamefont{and} \bibinfo{author}{\bibfnamefont{K.~H.} \bibnamefont{Ploog}},
  \bibinfo{journal}{Phys.\ Rev.\ Lett.} \textbf{\bibinfo{volume}{87}},
  \bibinfo{pages}{016601} (\bibinfo{year}{2001}).

\bibitem[{\citenamefont{Jiang et~al.}(2003)\citenamefont{Jiang, Wang, van
  Dijken, Shellby, Macfarlane, Solomon, Harris, and Parkin}}]{jwd03}
\bibinfo{author}{\bibfnamefont{X.}~\bibnamefont{Jiang}},
  \bibinfo{author}{\bibfnamefont{R.}~\bibnamefont{Wang}},
  \bibinfo{author}{\bibfnamefont{S.}~\bibnamefont{van Dijken}},
  \bibinfo{author}{\bibfnamefont{R.}~\bibnamefont{Shellby}},
  \bibinfo{author}{\bibfnamefont{R.}~\bibnamefont{Macfarlane}},
  \bibinfo{author}{\bibfnamefont{G.~S.} \bibnamefont{Solomon}},
  \bibinfo{author}{\bibfnamefont{J.}~\bibnamefont{Harris}}, \bibnamefont{and}
  \bibinfo{author}{\bibfnamefont{S.~S.~P.} \bibnamefont{Parkin}},
  \bibinfo{journal}{Phys.\ Rev.\ Lett.} \textbf{\bibinfo{volume}{90}},
  \bibinfo{pages}{256603} (\bibinfo{year}{2003}).

\bibitem[{\citenamefont{Datta and Das}(1990)}]{Datta:90.1}
\bibinfo{author}{\bibfnamefont{S.}~\bibnamefont{Datta}} \bibnamefont{and}
  \bibinfo{author}{\bibfnamefont{B.}~\bibnamefont{Das}},
  \bibinfo{journal}{Appl.~Phys.~Lett.} \textbf{\bibinfo{volume}{56}},
  \bibinfo{pages}{665} (\bibinfo{year}{1990}).

\bibitem[{\citenamefont{LaShell et~al.}(1996)\citenamefont{LaShell, McDougall,
  and Jensen}}]{LaShell:96.1}
\bibinfo{author}{\bibfnamefont{S.}~\bibnamefont{LaShell}},
  \bibinfo{author}{\bibfnamefont{B.~A.} \bibnamefont{McDougall}},
  \bibnamefont{and} \bibinfo{author}{\bibfnamefont{E.}~\bibnamefont{Jensen}},
  \bibinfo{journal}{Phys.\ Rev.\ Lett.} \textbf{\bibinfo{volume}{77}},
  \bibinfo{pages}{3419} (\bibinfo{year}{1996}).

\bibitem[{\citenamefont{Reinert et~al.}(2001)\citenamefont{Reinert, Nicolay,
  Schmidt, Ehm, and H\"uf\-ner}}]{Reinert:01.1}
\bibinfo{author}{\bibfnamefont{F.}~\bibnamefont{Reinert}},
  \bibinfo{author}{\bibfnamefont{G.}~\bibnamefont{Nicolay}},
  \bibinfo{author}{\bibfnamefont{S.}~\bibnamefont{Schmidt}},
  \bibinfo{author}{\bibfnamefont{D.}~\bibnamefont{Ehm}}, \bibnamefont{and}
  \bibinfo{author}{\bibfnamefont{S.}~\bibnamefont{H\"uf\-ner}},
  \bibinfo{journal}{Phys.\ Rev.\ B} \textbf{\bibinfo{volume}{63}},
  \bibinfo{pages}{115415} (\bibinfo{year}{2001}).

\bibitem[{\citenamefont{Rotenberg and Kevan}(1998)}]{Rotenberg:98.1}
\bibinfo{author}{\bibfnamefont{E.}~\bibnamefont{Rotenberg}} \bibnamefont{and}
  \bibinfo{author}{\bibfnamefont{S.~D.} \bibnamefont{Kevan}},
  \bibinfo{journal}{Phys.\ Rev.\ Lett.} \textbf{\bibinfo{volume}{80}},
  \bibinfo{pages}{2905} (\bibinfo{year}{1998}).

\bibitem[{\citenamefont{Hochstrasser et~al.}(2002)\citenamefont{Hochstrasser,
  Tobin, Rotenberg, and Kevan}}]{Rotenberg:02.1}
\bibinfo{author}{\bibfnamefont{M.}~\bibnamefont{Hochstrasser}},
  \bibinfo{author}{\bibfnamefont{J.~G.} \bibnamefont{Tobin}},
  \bibinfo{author}{\bibfnamefont{E.}~\bibnamefont{Rotenberg}},
  \bibnamefont{and} \bibinfo{author}{\bibfnamefont{S.~D.} \bibnamefont{Kevan}},
  \bibinfo{journal}{Phys.\ Rev.\ Lett.} \textbf{\bibinfo{volume}{89}},
  \bibinfo{pages}{216802} (\bibinfo{year}{2002}).

\bibitem[{\citenamefont{Nicolay et~al.}(2001)\citenamefont{Nicolay, Reinert,
  H\"ufner, and Blaha}}]{Nicolay:01.1}
\bibinfo{author}{\bibfnamefont{G.}~\bibnamefont{Nicolay}},
  \bibinfo{author}{\bibfnamefont{F.}~\bibnamefont{Reinert}},
  \bibinfo{author}{\bibfnamefont{S.}~\bibnamefont{H\"ufner}}, \bibnamefont{and}
  \bibinfo{author}{\bibfnamefont{P.}~\bibnamefont{Blaha}},
  \bibinfo{journal}{Phys.\ Rev.\ B} \textbf{\bibinfo{volume}{65}},
  \bibinfo{pages}{033407} (\bibinfo{year}{2001}).

\bibitem[{\citenamefont{Henk et~al.}(2003)\citenamefont{Henk, Ernst, and
  Bruno}}]{Henk:03.1}
\bibinfo{author}{\bibfnamefont{J.}~\bibnamefont{Henk}},
  \bibinfo{author}{\bibfnamefont{A.}~\bibnamefont{Ernst}}, \bibnamefont{and}
  \bibinfo{author}{\bibfnamefont{P.}~\bibnamefont{Bruno}},
  \bibinfo{journal}{Phys.\ Rev.\ B} \textbf{\bibinfo{volume}{68}},
  \bibinfo{pages}{165416} (\bibinfo{year}{2003}).

\bibitem[{\citenamefont{Weschke et~al.}(1996)\citenamefont{Weschke,
  Sch\"ussler-Langeheine, Meier, Fedorov, Starke, H\"ubinger, and
  Kaindl}}]{Weschke:96.1}
\bibinfo{author}{\bibfnamefont{E.}~\bibnamefont{Weschke}},
  \bibinfo{author}{\bibfnamefont{C.}~\bibnamefont{Sch\"ussler-Langeheine}},
  \bibinfo{author}{\bibfnamefont{R.}~\bibnamefont{Meier}},
  \bibinfo{author}{\bibfnamefont{A.~V.} \bibnamefont{Fedorov}},
  \bibinfo{author}{\bibfnamefont{K.}~\bibnamefont{Starke}},
  \bibinfo{author}{\bibfnamefont{F.}~\bibnamefont{H\"ubinger}},
  \bibnamefont{and} \bibinfo{author}{\bibfnamefont{G.}~\bibnamefont{Kaindl}},
  \bibinfo{journal}{Phys.\ Rev.\ Lett.} \textbf{\bibinfo{volume}{77}},
  \bibinfo{pages}{3415} (\bibinfo{year}{1996}).

\bibitem[{\citenamefont{Berger et~al.}(1994)\citenamefont{Berger, Pang, and
  Hop\-ster}}]{Berger:94.1}
\bibinfo{author}{\bibfnamefont{A.}~\bibnamefont{Berger}},
  \bibinfo{author}{\bibfnamefont{A.~W.} \bibnamefont{Pang}}, \bibnamefont{and}
  \bibinfo{author}{\bibfnamefont{H.}~\bibnamefont{Hop\-ster}},
  \bibinfo{journal}{J.\ Magn.\ Magn.\ Mater.} \textbf{\bibinfo{volume}{137}},
  \bibinfo{pages}{L1} (\bibinfo{year}{1994}).

\bibitem[{\citenamefont{Sch\"ussler-Langeheine
  et~al.}(1999)\citenamefont{Sch\"ussler-Langeheine, Meier, Ott, Hu, Mazumdar,
  Grigoriev, Kaindl, and Weschke}}]{Schussler:99.1}
\bibinfo{author}{\bibfnamefont{C.}~\bibnamefont{Sch\"ussler-Langeheine}},
  \bibinfo{author}{\bibfnamefont{R.}~\bibnamefont{Meier}},
  \bibinfo{author}{\bibfnamefont{H.}~\bibnamefont{Ott}},
  \bibinfo{author}{\bibfnamefont{Z.}~\bibnamefont{Hu}},
  \bibinfo{author}{\bibfnamefont{C.}~\bibnamefont{Mazumdar}},
  \bibinfo{author}{\bibfnamefont{A.~Y.} \bibnamefont{Grigoriev}},
  \bibinfo{author}{\bibfnamefont{G.}~\bibnamefont{Kaindl}}, \bibnamefont{and}
  \bibinfo{author}{\bibfnamefont{E.}~\bibnamefont{Weschke}},
  \bibinfo{journal}{Phys.\ Rev.\ B} \textbf{\bibinfo{volume}{60}},
  \bibinfo{pages}{3449} (\bibinfo{year}{1999}).

\bibitem[{\citenamefont{Moruzzi et~al.}(1978)\citenamefont{Moruzzi, Janak, and
  Williams}}]{Moruzzi:78.1}
\bibinfo{author}{\bibfnamefont{V.}~\bibnamefont{Moruzzi}},
  \bibinfo{author}{\bibfnamefont{J.}~\bibnamefont{Janak}}, \bibnamefont{and}
  \bibinfo{author}{\bibfnamefont{A.}~\bibnamefont{Williams}},
  \emph{\bibinfo{title}{Calculated Electronic Properties of Metals}}
  (\bibinfo{publisher}{Pergamon}, \bibinfo{address}{New York},
  \bibinfo{year}{1978}).

\bibitem[{\citenamefont{Wimmer et~al.}(1981)\citenamefont{Wimmer, Krakauer,
  Weinert, and Freeman}}]{Wimmer:81.1}
\bibinfo{author}{\bibfnamefont{E.}~\bibnamefont{Wimmer}},
  \bibinfo{author}{\bibfnamefont{H.}~\bibnamefont{Krakauer}},
  \bibinfo{author}{\bibfnamefont{M.}~\bibnamefont{Weinert}}, \bibnamefont{and}
  \bibinfo{author}{\bibfnamefont{A.}~\bibnamefont{Freeman}},
  \bibinfo{journal}{Phys.\ Rev.\ B} \textbf{\bibinfo{volume}{24}},
  \bibinfo{pages}{864} (\bibinfo{year}{1981}).

\bibitem[{www()}]{www-fleur}
\bibinfo{howpublished}{For program description see {\em http://www.flapw.de}.}

\bibitem[{\citenamefont{Li et~al.}(1990)\citenamefont{Li, Freeman, Jansen, and
  Fu}}]{Li:90.1}
\bibinfo{author}{\bibfnamefont{C.}~\bibnamefont{Li}},
  \bibinfo{author}{\bibfnamefont{A.~J.} \bibnamefont{Freeman}},
  \bibinfo{author}{\bibfnamefont{H.~J.~F.} \bibnamefont{Jansen}},
  \bibnamefont{and} \bibinfo{author}{\bibfnamefont{C.~L.} \bibnamefont{Fu}},
  \bibinfo{journal}{Phys.\ Rev.\ B} \textbf{\bibinfo{volume}{49}},
  \bibinfo{pages}{5433} (\bibinfo{year}{1990}).

\bibitem[{\citenamefont{Kurz et~al.}(2002)\citenamefont{Kurz, Bihl\-mayer, and
  Bl\"ugel}}]{Kurz:02.1}
\bibinfo{author}{\bibfnamefont{P.}~\bibnamefont{Kurz}},
  \bibinfo{author}{\bibfnamefont{G.}~\bibnamefont{Bihl\-mayer}},
  \bibnamefont{and} \bibinfo{author}{\bibfnamefont{S.}~\bibnamefont{Bl\"ugel}},
  \bibinfo{journal}{J.~Phys.: Cond. Matter} \textbf{\bibinfo{volume}{14}},
  \bibinfo{pages}{6353} (\bibinfo{year}{2002}).

\end{thebibliography}

\end{document}